\documentclass[journal, 12]{IEEEtran}
\usepackage[pdftex]{graphicx}
\graphicspath{{../pdf/}{../jpeg/}}
\DeclareGraphicsExtensions{.pdf,.jpeg,.png}
 \usepackage[cmex10]{amsmath}
\usepackage{amsfonts}
\usepackage[english]{babel}
\usepackage{amssymb}

\usepackage{graphicx,color}
\usepackage{cite}
\allowdisplaybreaks
\usepackage{blindtext}
\usepackage{tcolorbox}
\usepackage{xcolor}
\usepackage{lipsum}
\usepackage[font={small}]{caption}
\usepackage{algorithm}
\usepackage{algorithmic}
\usepackage{array} 
\usepackage{makecell}
\usepackage{tabularx}
\usepackage[font={small}]{caption}
\usepackage{subcaption}
\usepackage{enumerate}
\usepackage{float} 
\def\bUpsilon{\boldsymbol{\Upsilon}}

\def\bI{\mathbf{I}}
\def\br{\mathbf{r}}
\def\bQ{\mathbf{T}}

\def\bP{\mathbf{P}}

\def\bu{\mathbf{u}}

\begin{document}

\title{ Multiuser MISO PS-SWIPT Systems: Active or Passive RIS? }

\author{ Shayan Zargari, Azar Hakimi, and Chintha Tellambura, \IEEEmembership{Fellow, IEEE,}   Sanjeewa Herath, \IEEEmembership{Member, IEEE}
\thanks{S. Zargari, A. Hakimi, and C. Tellambura with the Department of Electrical and Computer Engineering, University of Alberta, Edmonton, AB, T6G 1H9, Canada (e-mail: {zargari,hakimina,ct4}@ualberta.ca).  \\
\indent S. Herath is with the Huawei Canada, 303 Terry Fox Drive, Suite 400, Ottawa, Ontario K2K 3J1 (e-mail: sanjeewa.herath@huawei.com).} \vspace{-5mm}  }

\maketitle

\begin{abstract}
Reconfigurable intelligent surface (RIS)-based communication networks promise to improve channel capacity and energy efficiency. However,  the promised capacity gains could be negligible for passive RISs because of the double pathloss effect. Active RISs can overcome this issue because they have reflector elements with a low-cost amplifier. This letter studies the active RIS-aided simultaneous wireless information and power transfer (SWIPT) in a multiuser system. The users exploit power splitting (PS) to decode information and harvest energy simultaneously based on a realistic piecewise nonlinear energy harvesting model. The goal is to minimize the base station (BS) transmit power by optimizing its beamformers, PS ratios, and RIS phase shifts/amplification factors. The simulation results show significant improvements (e.g., $ 19 \% $ and $28 \%$) with the maximum reflect power of $10$ mW and $15$ mW, respectively, compared to the passive RIS without higher computational complexity cost. We also show the robustness of the proposed algorithm against imperfect channel state information.

\end{abstract}


\vspace{-3mm}

\section{Introduction}
A reconfigurable intelligent surface (RIS), a metasurface with numerous low-cost reflectors, can inherently manipulate the propagation environment to enhance energy and spectrum efficiency~\cite{renzo2020,zhang2021}. RIS can be passive or active. The former imparts a phase change only, whereas the latter changes both the phase and amplitude of the incident radio frequency  (RF) signal. Both types avoid expensive RF components, which are highly cost-effective. Thus,   RIS may enable wireless fidelity (WiFi), navigation, and a plethora of applications \cite{zhang2021}.  Passive RIS and energy harvesting have also been considered \cite{Shaokang}.  However, the double pathloss from the product of the base station (BS)-RIS channel and the  RIS-user channel limits the performance of the passive RIS  networks  \cite{zhang2021}.  Increasing the number of reflecting may mitigate this loss but drives up the cost. In response, active RIS has emerged   \cite{long2021,zhang2021,xu2021,YouZ21}. The active RIS  includes a  negative resistance so that the reflectors can not only reflect the RF signal but amplify it.   

The benefits of active RISs have already been investigated \cite{long2021,xu2021}.
Reference  \cite{long2021} compares the active and passive RIS in a single-input multi-output (MISO)  network and formulates the signal-to-noise ratios (SNRs).  The spectral efficiency of the uplink is investigated in \cite{JungSK21}, where asymptotic and theoretical performance bounds are derived.  \cite{xu2021} develops resource allocations for a multiuser network by minimizing the BS transmit power. Indoor and outdoor applications can deploy active RIS-aided systems with energy harvesting (EH) requirements,  such as smart homes and IoT networks. This deployment improves the network energy efficiency, a critical benefit. On the other, simultaneous wireless information and power transfer (SWIPT)  can help the same goals  \cite{Zargari1,Shayan_Zargari_1,Shayan_Zargari_2}. 
 
This letter aims to quantify the benefits of active RIS by evaluating transmit power savings achieved by a network resulting from its use. Accordingly, we design the system with an active RIS (Fig. 1) to minimize the transmit power of the BS. This optimization problem involves the BS beamformers, power splitting (PS) ratios at the users, and active RIS phase shifts/amplification factors. This problem is non-convex because of variable entanglement. We thus leverage the block coordinate descent (BCD) method (also known as alternating optimization), which divides the variables into two alternating blocks. The first block comprises the BS beamformer weights and the PS ratios,  and the second block comprises the active RIS phase shifts and amplification factors. The methods we use to solve them are convex relaxation and a penalty-based technique. We thus develop the intermediate power and phase shift algorithm (IPPA) and the overall BCD algorithm. We also show that performance gain for the active RIS over the passive RIS comes with no additional computational complexity cost.

\textit{Notation:} Vectors and matrices are indicated by boldface lower-case  and capital letters, respectively. For a square matrix $\mathbf{A}$, $\mathbf{A}^H$, $\mathbf{A}^T$, $\text{Tr}(\mathbf{A})$, $||\mathbf{A}||_{*}$, and $\text{Rank}(\mathbf{A})$ denote its Hermitian conjugate transpose, transpose, trace, trace norm, and Rank, respectively. $\mathbf{A}\succeq\mathbf{0}$ denotes  a positive semidefinite matrix. $\text{diag}(\cdot)$ is the diagonalization operation. The Euclidean norm of $\mathbf{x}$ is  $\|\mathbf x \|,$ and the absolute value of $x$ is  $|x|$.  $\nabla_{\mathbf{x}}f(\mathbf{x})$ is the  gradient vector over $\mathbf{x}$.  $\mathbb{E}[x]$ is the expectation of $x$. A circularly symmetric complex Gaussian (CSCG) random vector with mean $\boldsymbol{\mu}$ and covariance matrix $\mathbf{C}$ is denoted by $\sim \mathcal{C}\mathcal{M}(\boldsymbol{\mu},\,\mathbf{C})$. $\mathbb{C}^{M\times N}$ indicates $M\times N$ dimensional complex matrices. $\mathcal{O}$ denotes the big-O notation.

 \vspace{-3mm}
 
%%%%%%%%%%%%%%%%%%%%%%%%%%%%%%%%%%%%%%%%%%%%%%
                %SYSTEM MODEL%
%%%%%%%%%%%%%%%%%%%%%%%%%%%%%%%%%%%%%%%%%%%%%% 
\section{System Model}\label{system}
 
	We consider a downlink communication system with a $M$ antennas  BS and $K$ single antenna users index by $\mathcal{K}=\{1,\ldots, K \}$. We propose an active RIS with $N$ reflecting elements to improve this communication link. At the users,  the received signal is split for information decoding (ID) and EH. The BS uses each time-frequency resource block simultaneously for transmitting to all the users. The flat-fading channel gains from the BS-to-RIS, RIS-to-user $k$, and BS-to-user $k$ are denoted by ${\mathbf{G}}\in \mathbb{C}^{N \times M}$, ${\mathbf{h}}_{r,k}\in \mathbb{C}^{N\times 1}$, and ${\mathbf{h}}_{b,k}\in \mathbb{C}^{M\times 1}$,~respectively.  All the channels undergo quasi-static flat Rician fading and remain unchanged for several symbols  \cite{Shayan_Zargari_1}. We also assume the  RIS controller acquires all the channel state information (CSI) during the channel estimation phase\footnote{If perfect CSI is not available, our results serve as theoretical performance upper bounds for the considered system. Robust versions of our algorithm can also be developed to reduce the effect of channel estimation errors, e.g., \cite{Shayan_Zargari_2}.}. The system then adopts a codebook-based passive beamforming approach, where the RIS selects the best beam based on the received pilot sequences from the BS or users \cite{renzo2020}. %Alternatively, if CSI is unavailable,  non-coherent detection techniques can be adapted for each user to decode information. %\cite{tse2005fundamentals}.
	%\subsection{Proposed Active RISs}
	
	 The signal model of  $N$ elements passive RIS is given by $
	    \br=\mathbf{\Theta}\mathbf{x}$, where ${ {\mathbf{\Theta}}=\text{diag} (e^{j\theta_{1}},\ldots ,e^{j\theta_{N}})}$ is the reflection-coefficients matrix at the RIS.
	    %, and the effect of noise is neglected. 
	    With optimal  RIS phase shifts,  coherent combining at the user yields a  high array gain proportional to $N^2,$  a fundamental motivation of employing RIS \cite{renzo2020}. However, when the direct link between the BS and the user is strong enough, the array gain becomes marginal \cite{zhang2021}. This loss of efficacy happens because of the double pathloss effect, the product term of the BS-RIS channel and RIS-user channel. For a network at  5 GHz,  $10,000$ RIS elements are required to make the reflection link as strong as the direct link \cite{zhang2021}. We thus consider an active RIS  to overcome this issue. Unlike passive RIS, the active RIS can amplify the reflected signals\footnote{To realize the active RIS platform, we assume an active reflection-type amplifier, which active elements can equip, e.g., current-inverting converters, asymmetric current mirrors, or some integrated circuits  \cite{zhang2021}}. Unlike full-duplex amplify-and-forward (FD-AF) relays, active RISs lack RF components and signal processing capabilities but only reflect and amplify the incident signals. Also, the performance of a FD-AF is limited by self-interference. Thus, a detailed comparison between these is an important future topic. Accordingly, the amplified  and reflected signal of the active RIS \cite{zhang2021, xu2021} can be represented as  $ 
	        \br=\bP\mathbf{\Theta}\mathbf{x}+\bP\mathbf{\Theta}\mathbf{v}+\boldsymbol{\nu} , $ where $\bP=\text{diag} (p_1,\ldots ,p_N)\in \mathbb{R}^{N\times N}$  denotes the amplification factor matrix of the active RIS with each component greater than one. The second and third terms are dynamic and static noise at the active RIS. In addition, $\mathbf{v}\sim \mathcal{C}\mathcal{N}(\mathbf{0}_N,\sigma_{v}^2\bI_N)$ is related to the input noise and the inherent device noise of the active RIS elements, while the static noise $\boldsymbol{\nu}$ is unrelated to $\bP$ and is negligible \cite{zhang2021,xu2021}.

	\section{System Model}
	\begin{figure}[t]
		\centering
		\includegraphics[width=3in]{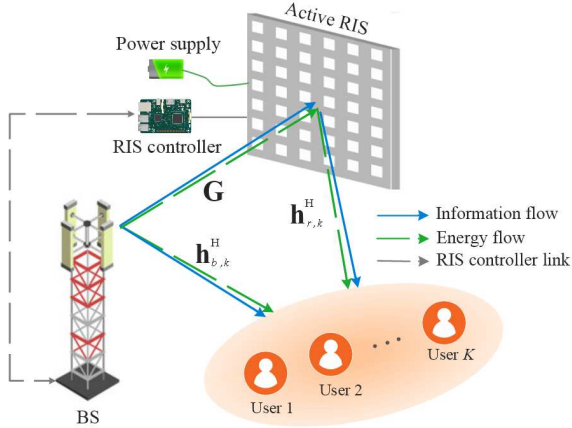}
		\caption{\small {A multiuser MISO active RIS-assisted SWIPT system.}}
	\end{figure}

%	\subsection{transmit Scheme}
	The transmitted signal by the BS is given by $\mathbf{x} =\sum_{i = 1}^K {{\mathbf{w}_i}{s_i}}$, where ${\mathbf{w}}_{i}\in \mathbb{C}^{M \times 1}$, $\forall i\in \mathcal{K},$ denotes the BS beamformers and $s_{i}$ is the information symbol intended for user $i$ which satisfies $\mathbb{E}\left[|s_{i}|^2\right]=1,~i\in\mathcal{K}$. %It is assumed that $s_i$, $\forall i\in \mathcal{K},$ are independent random variables with zero-mean and unit-variance. 
	Then, the received signal at the active RIS is given by  $
	\tilde{\br} = \bP \mathbf{\Theta}\mathbf{G} \sum_{i = 1}^K  {{\mathbf{w}_i}{s_i}} +\bP\mathbf{\Theta}\mathbf{v}.$
	Subsequently, the received signal at user $k$ becomes $
	{y_k} = \sum_{i = 1}^K \mathbf{h}_k^H{{\mathbf{w}_i}{s_i}}+ \mathbf{h}_{r,k}^H\bP\mathbf{\Theta}\mathbf{v} + {z_k},~\forall k , $ 
	where $\mathbf{h}^H_k=\mathbf{h}_{r,k}^H\bP\mathbf{\Theta}\mathbf{G}+{\mathbf{h}}_{b,k}^H$ denotes the total channel gain from the BS to user $k$. Also, $z_{k}\sim \mathcal{C}\mathcal{N}(0,\sigma_{k}^2)$ is the received complex Gaussian noise at user $k$. By denoting $\rho_k \in (0,1)$ as the PS ratio, the received signal at each user is divided into two portions, where $\rho_k$ portion is used for ID, and $(1-\rho_k)$ remaining portion is utilized for the EH. Consequently, the received signals for the ID and EH sections are given by $y_{k}^{\text{ID}}=\sqrt{\rho_{k}}y_k+n_{k}$ and $
	y_{k}^{\text{EH}}=\sqrt{1-\rho_{k}}y_k$, respectively, where $n_{k}\sim \mathcal{C}\mathcal{N}(0,\,\delta_{k}^2)$ is the additional noise introduced by the signal processing circuit of user $k$ for the ID section \cite{Shayan_Zargari_1}. Hence, the received signal-to-interference-plus-noise ratio (SINR) at  user $k$ can be written as
	\begin{equation}\label{sinr}
	\hspace{-0.7mm}\text{SINR}_k	\!=  \!\frac{{{{\left| {{{\rm{ {\mathbf{h}}}}}}_{k}^H {\mathbf{{{w}}}}_k \right|}^2}}}{{\sum\limits_{\scriptstyle i = 1\atop\scriptstyle i \ne k}^K {{{\left|  {{{\rm{ {\mathbf{h}}}}}}_{k}^H {\mathbf{{{w}}}}_i \right|}^2}+\sigma^2_v\|\mathbf{h}_{r,k}^H\bP\mathbf{\Theta}\|^2 +\sigma _k^2 + \frac{\delta _k^2}{{\rho _k}}} }},~\forall k .
	\end{equation}
	 
%	\subsection{{Non-linear Energy Harvesting Model}}
		Practical EH circuits have two major drawbacks. First, the harvested energy does not linearly increase when the input power increases; instead, it saturates. Second, when the input power drops below the sensitivity level of the EH circuit, the harvested energy is zero \cite{Shayan_Zargari_1,Shayan_Zargari_2}. To capture these effects precisely, we employ the following piecewise nonlinear EH model \cite{Yunfei_Chen}:   $P_\text{EH}^{\text{NL}}= \frac{aP_k^{\text{L}}+b}{P_k^{\text{L}}+c}-\frac{b}{c}$, where  $P_k^{\text{L}}$ represents the linear input  power to the EH section  and $a > 0$, $b> 0,$ and $c > 0$ capture the saturation and sensitivity thresholds of the EH circuit. By applying a best-fit match with experimental data, $a$, $b$, and $c$ can be estimated. The  harvested energy at   user $k$ can then  be written as $
	P_k^{\text{L}}=\eta_k(1-\rho_{k})\left( \sum\limits_{i = 1}^K {{\left|  {{{\rm{ {\mathbf{h}}}}}}_{k}^H {\mathbf{{{w}}}}_i \right|}^2}+  \sigma^2_v\|\mathbf{h}_{r,k}^H\bP\mathbf{\Theta}\|^2  \right),~\forall k ,$ 
	where $\eta_k\in(0, 1]$ is the energy conversion efficiency. We assume $\eta_k=1$ for all users.  

	\section{Optimization Problem}
	We aim to minimize the BS transmit power\footnote{To realize the potential gain due to the use of an  active RIS, a proper amount of power is allocated to each active RIS element from the total budget. Thus, compared to the passive RIS, the precise design of the BS beamformers is more important such that the quality of service for each user can be fulfilled, and the system power consumption is not excessive  \cite{xu2021}.}  by jointly optimizing the BS beamformers, PS ratios, and RIS phase shifts/amplification factors. The optimization problem can be formulated as
	\begin{subequations}
		\begin{align}
		\text{(P1)}\!: \:	& \underset{  {\mathbf{w}}_k ,{\mathbf{\Theta}},\bP,\rho_k} {\text{minimize}} \quad  f_1 = \sum_{k=1}^{K}{\left\| {\mathbf{w}}_k \right\|_2^2},\\
		&\text{s.t.} \:\: \frac{{{{\left| {{{\rm{ {\mathbf{h}}}}}}_{k}^H {\mathbf{{{w}}}}_k \right|}^2}}}{{\sum\limits_{\scriptstyle i = 1\atop\scriptstyle i \ne k}^K\! {{{\left|  {{{\rm{ {\mathbf{h}}}}}}_{k}^H {\mathbf{{{w}}}}_i \right|}^2}\!+\!\sigma^2_v\|\mathbf{h}_{r,k}^H\bP\mathbf{\Theta}\|^2 \!+\!\sigma _k^2\! +\! \frac{\delta _k^2}{{\rho _k}}} }}\!\geq\! \gamma_{k}, \label{p1-1}\\
		&\:\:\quad \sum\limits_{i = 1}^K \|\bP \mathbf{\Theta}\mathbf{G}\mathbf{w}_i\|^2+\sigma^2_v\| \bP\mathbf{\Theta}\|^2\leq p_{\max}  , \label{p1-3}\\
		&\:\:\quad 	P_\text{EH}^{\text{NL}}\geq  e_{k} , \: \forall k ,\label{p1-2}\\
		&\:\:\quad 0 < \rho_{k}< 1, \: \forall k ,\label{p1-4}
		\end{align}
	\end{subequations}
	where $\gamma_{k}$ and ${e}_{k}$ in (\ref{p1-1}) and (\ref{p1-2}) are the minimum SINR and harvested energy requirements at user $k$, respectively. Constraint \eqref{p1-3} indicates the power constraints at the active RIS with maximum reflect power, $p_{\max}$.
	Constraint (\ref{p1-4}) denotes the PS ratio requirements. Following the similar approach in \cite{Shayan_Zargari_1}, we rewrite the right-hand side of constraint (\ref{p1-2}) in a tractable form as follows: $
	P_k^{\text{L}}(P_\text{EH}^{\text{NL}})=\frac{b-(P_\text{EH}^{\text{NL}}+\frac{b}{c})c}{P_\text{EH}^{\text{NL}}+\frac{b}{c}-a}$. Accordingly, (P1) can be recast as below
	\begin{subequations}
		\begin{align}
		\text{(P2)}:\:\:	&  \underset{ {\mathbf{w}}_k ,{\mathbf{\Theta}},\bP,\rho_k} {\text{minimize}} \quad f_1 = \sum_{k=1}^{K}{\left\| {\mathbf{w}}_k \right\|_2^2},\\
		&\text{s.t.}  \:\: \text{ (\ref{p1-1}), (\ref{p1-3}), (\ref{p1-4})},\\
		&\:\:\quad \sum\limits_{i = 1}^K {{\left|  {{{\rm{ {\mathbf{h}}}}}}_{k}^H {\mathbf{{{w}}}}_i \right|}^2}+  \sigma^2_v\|\mathbf{h}_{r,k}^H\bP\mathbf{\Theta}\|^2 \geq \frac{P_k^{\text{L}}(e_{k})}{	1-\rho_{k}},\label{P1-5c}
		\end{align}
	\end{subequations}
(P2) is non-convex as the constraints contain entangled terms. Thus, we cannot solve it by standard popular convex software, e.g., CVX  \cite{18}. Therefore, we use a divide-and-conquer method, namely the BCD algorithm. First, we optimize the BS beamformers and PS ratios by applying semidefinite relaxation (SDR) and successive convex approximation (SCA). Second, we design the RIS phase shifts and amplification factors given the output of the first part. This process iterates until convergence.

	\subsection{{Optimizing $\mathbf{w}_k$ and $\rho_k$  With Given $\mathbf{\Theta}$} and $\bP$ }
		Since  $\bP$  and $\mathbf{\Theta}$ appears in product form, we equivalently optimize $\bUpsilon= \bP {\mathbf{\Theta}}=\text{diag} (p_1 e^{j\theta_{1}},\ldots ,p_Ne^{j\theta_{N}}) $ as the RIS precoding matrix that facilitates the optimization problem. By defining $\mathbf{W}_k=\mathbf{w}_k\mathbf{w}^H_k$ and $\mathbf{H}_k=\mathbf{h}_k\mathbf{h}^H_k$ $\forall k $, we have
	\begin{subequations}
		\begin{align}
	\hspace{-3mm}	\text{(P3)}\!: \:	& \underset{ \mathbf{W}_k,\rho_k} {\text{minimize}} \quad f_2 =\sum_{k=1}^{K} \text{Tr}({\mathbf{W}}_k) ,\\
		&\!\text {s.t.} \:\:\! \frac{{\text{Tr}({{\mathbf{H}}_k}\!{{\mathbf{W}}_k})}}{{{\gamma _k}}} \!\!-\!\!\sum\limits_{\scriptstyle i = 1\atop\scriptstyle i \ne k}^K\! {\text{Tr}({{\mathbf{H}}_k}\!{{\mathbf{W}}_i})} \!\ge\! \sigma^2_v\|\mathbf{h}_{r,k}^H\bUpsilon\|^2\!+\! \tilde{\sigma}_k, \label{P2-1}\\
		&\:\:\quad {\sum\limits_{i = 1}^K 
		\text{Tr}( \bUpsilon \mathbf{G} \mathbf{W}_i \mathbf{G}^H \bUpsilon^H)+\sigma^2_v\| \bUpsilon\|^2\leq p_{\max}},\\
		&\:\:\quad \sum\limits_{i = 1}^K {{ \text{Tr}({{\mathbf{H}}_k}{{\mathbf{W}}_i}) }+\sigma^2_v\|\mathbf{h}_{r,k}^H\bUpsilon\|^2  \ge \frac{{{P_k^{\text{L}}(e_{k})}}}{{1 - {\rho _k}}} }, \label{P2-2}\\
		&\:\:\quad  0 \le \rho_{k} \le 1,\quad {\mathbf{W}}_k\succeq 0,~\forall k ,\label{P3-psd}
		\end{align} 
	\end{subequations}
	where $\tilde{\sigma}_k :=\sigma _k^2 + \frac{{\delta _k^2}}{{{\rho _k}}}$, $\forall k\in \mathcal{K}$. The rank-one constraint, i.e., ${\text{Rank}(\mathbf{W}_k)\leq 1}$,~$\forall k ,$ is omitted to make (P3) a convex problem. (P3) is a standard semidefinite program (SDP)  that can be solved efficiently by using  CVX \cite{18}. For all ${\bf W}_k$'s, {it can be shown that the optimal solution to (P3) satisfies rank-one constraints} \cite{Zargari1}.  Thus, eigenvalue decomposition (EVD) yields a globally optimal solution.  The computational complexity of (P3) is given by $\mathcal{O}\left( \sqrt{KM }(K^3M^2 + K^2M^3)\right)$ \cite{Zargari1}. 

	\subsection{{Optimizing $\bP$  and $\mathbf{\Theta}$   with Given $\mathbf{w}_k$ and $\rho_k$ }}
	%Subsequently, the RIS phase shifts/amplication factors are optimized in this subsection by fixing $\{\bw_k,\rho_k\}$ in the previous subproblem. 
 To design the RIS precoding matrix, we transform the optimization problem  into a feasibility check problem. 	
	Let us first define ${\boldsymbol{\theta}} =(p_1 e^{j\theta_{1}},\ldots ,p_Ne^{j\theta_{N}})^H\in\mathbb{C}^{N\times1}$. Furthermore, to make the problem more tractable, we apply the change of variables ${\mathbf{h}}_{b,k}^\text{H}{\mathbf{w}}_i = {a_{k,i}}$,    ${\mathbf{h}}_{r,k}^\text{H}{\bUpsilon} {\mathbf{G}}{\mathbf{w}}_i  = {{\boldsymbol{\theta}}^\text{H}}{{\mathbf{b}}_{k,i}}$, where ${{\mathbf{b}}_{k,i}} = \text{diag}({\mathbf{h}}_{r,k}^\text{H}){\mathbf{G}}{\mathbf{w}}_i $, $\mathbf{Q}_k=  \text{diag}(\mathbf{G}{\mathbf{w}}_k)(\text{diag}(\mathbf{G}{\mathbf{w}}_k))^H $,  and $\mathbf{Z}_k=  \text{diag}(\mathbf{h}_{r,k}^\text{H})\text{diag}(\mathbf{h}_{r,k} ) $, $\forall k \in \mathcal{K}$. Then, (P1) can be rewritten as follows
\begin{subequations}
	\begin{align}
	(\text{P4})\!:\:\:&\text{Find}\quad  \boldsymbol{\theta},\\
	&\text{s.t.} \: \: \frac{{{{\left| {{{\boldsymbol{\theta}}^\text{H}}{{\mathbf{b}}_{k,k}} + {a_{k,k}}} \right|}^2}}}{{\sum\limits_{\scriptstyle j = 1\atop\scriptstyle j \ne k}^K {{{\left| {{{{\boldsymbol{\theta}}^\text{H}}}{{\mathbf{b}}_{k,j}} + {a_{k,j}}} \right|}^2} + \sigma^2_v{\boldsymbol{\theta}}^\text{H}\mathbf{Z}_k{\boldsymbol{\theta}}+\tilde{\sigma}_k} }}\geq \gamma_{k}, \label{p3-1}\\
	&\:\:\quad \sum\limits_{i = 1}^K {\boldsymbol{\theta}}^\text{H}\mathbf{Q}_k{\boldsymbol{\theta}}+\sigma^2_v{\boldsymbol{\theta}}^\text{H} {\boldsymbol{\theta}}\leq p_{\max},\\
	&\:\:\quad \sum\limits_{j = 1}^K  {{{\left| {{{\boldsymbol{\theta}}^\text{H}}{{\mathbf{b}}_{k,j}} + {a_{k,j}}} \right|}^2}}+\sigma^2_v{\boldsymbol{\theta}}^\text{H}\mathbf{Z}_k{\boldsymbol{\theta}} \geq \frac{{{P_k^{\text{L}}(e_{k})}}}{1-\rho_{k}}. \:
	\end{align}
\end{subequations}
Since  (P4) holds quadratic inequality and equality constraints, the whole problem is  non-convex. Hereafter, we use the SDR technique to transform  (P4) into a convex problem. We introduce $\tilde{{\boldsymbol{\theta}}}:=[{\boldsymbol{\theta}}^T \: 1]^T\in\mathbb{C}^{(N+1)\times1}$  and ${{\mathbf{S}}_{k,j}} = [{{\mathbf{b}}_{k,j}}{\mathbf{b}}_{k,j}^H,~{{\mathbf{b}}_{k,j}}a_{k,j}^{H};~{\mathbf{b}}_{k,j}^H{a_{k,j}},~0]$,~$\forall i\in \mathcal{K}$,~respectively. Further, we define ${\mathbf{T}}:=\tilde{{\boldsymbol{\theta}}}\tilde{{\boldsymbol{\theta}}}^H\in\mathbb{C}^{(N+1)\times(N+1)}$,   which requires to satisfy ${\mathbf{T}}\succeq\mathbf{0}$ and $\text{Rank}(\mathbf{T})=1$. Dropping the non-convex rank-one constraint, (P4) is relaxed into
\begin{subequations}
	\begin{align}
 \hspace{-2mm}	(\text{P5})\!:\:\:&\text{Find}  \quad \mathbf{T},\\
	&\text{s.t.} \:\: \hspace{1em} \frac{{\text{Tr}({{\mathbf{S}}_{k,k}}{\mathbf{T}})+|a_{k,k}|^2}}{{{\gamma _k}}} - \sum\limits_{\scriptstyle j = 1\atop\scriptstyle j \ne k}^K  \! {\text{Tr}({{\mathbf{S}}_{k,j}}{\mathbf{T}})}  \nonumber \\
	&\hspace{12em} \!- \!\sigma^2_v{\text{Tr}({{\tilde{\mathbf{Z}}}_k}{\mathbf{T}})}  \! \ge  \!\tilde{\sigma}_k,  \label{12b}\\
	&\:\:\quad \sum\limits_{i = 1}^K {\text{Tr}({{\tilde{\mathbf{Q}}}_k}{\mathbf{T}})} +\sigma^2_v{\text{Tr}( {\mathbf{T}})} \leq p_{\max},\\
	&\:\:\quad \sum\limits_{j = 1}^K {\text{Tr}({{\mathbf{S}}_{k,j}}{\mathbf{T}})+|a_{k,j}|^2}  +\sigma^2_v{\text{Tr}({{\tilde{\mathbf{Z}}}_k}{\mathbf{T}})}  \ge \frac{{{P_k^{\text{L}}(e_{k})}}}{{ 1 - {\rho _k}}},\label{12c}\\
	&\:\:\quad   {\mathbf{{{T}}}} \succeq 0,\label{kkk}
	\end{align}
\end{subequations}
where $\tilde{\mathbf{Q}}_k$ and $\tilde{\mathbf{Z}}_k$ are   matrices with extra zero rows and columns. (P5) is an SDP that can be solved using CVX \cite{18}. To achieve a rank-one solution, we exploit the penalty-based method \cite{Shayan_Zargari_1,Shayan_Zargari_2}. An equivalent form for the rank-one constraint can be written as
	$||\mathbf{T}||_*-||\mathbf{T}||_2\leq 0$. We know that the inequality $||\mathbf{T}||_*=\sum_{i} \sigma_i\geq ||\mathbf{T}||_2=\underset{i}{\text{max}}\{\sigma_i\}$ holds for any given $\mathbf{T}\in \mathbb{H}^{m\times n}$, where $\sigma_{i}$ is the $i$-th singular value of $\mathbf{T}$. The equality holds if and only if $\mathbf{T}$ has rank-one. We thus have   the following optimization problem:
% \begin{figure*}[t]
% \centering
% \begin{subfigure}[b]{0.32\textwidth}
% \includegraphics[width=\textwidth]{M.eps}
% \caption{BS transmit power versus\\ number of  antenna at the BS \\ with $N=20$.}   \label{antenna_fig}
% \end{subfigure}
% \begin{subfigure}[b]{0.32\textwidth}
% \includegraphics[width=\textwidth]{N.eps}
% \caption{BS transmit power versus   number\\ of   active elements at the RIS  with $M=10$.}      \label{phase_fig}
% \end{subfigure}
% \begin{subfigure}[b]{0.32\textwidth}
% \includegraphics[width=\textwidth]{csi.eps}
% \caption{BS transmit power versus   CSI\\ error parameter, $\xi$, with $M=10$.}   \label{channel_unc_fig}
% \end{subfigure}
% \begin{subfigure}[b]{0.32\textwidth}
% \includegraphics[width=\textwidth]{csi.eps}
% \caption{BS transmit power versus   CSI \\ error parameter, $\xi$, with $M=10$.}   \label{channel_unc_fig}
% \end{subfigure}
% \caption{ Performance analysis under different setups.}
% \end{figure*}	
\begin{algorithm}[t]
		\caption{ {Intermediate power and phase shift Algorithm  (IPPA)}}
		\begin{algorithmic}[1]
			\renewcommand{\algorithmicrequire}{\textbf{Input:}}
			\renewcommand{\algorithmicensure}{\textbf{Output:}}
			\REQUIRE Set the iteration number ${j}=0$, {the convergence tolerance $\zeta$},  and initialize $\mathbf{T}^{(0)}$.
			\STATE \textbf{repeat}\\
			\STATE \quad Calculate $\tilde{\Psi}(\mathbf{T})$ according to  (\ref{sca}). 
			\STATE \quad Solve   (P7) to obtain $\{\mathbf{T}^{{(j)}}\}$.\\
			\STATE \quad $j\leftarrow j+1$;
			\STATE  { \textbf{until} $\frac{|f_3^{(j)}-f_3^{(j-1)}|}{f_3^{(j-1)}}\geq \zeta$}
			\ENSURE return solution $\{\mathbf{T}^\ast\}$.
		\end{algorithmic}
	\end{algorithm}
	\begin{subequations}
		\begin{align}
	 	\text{(P6)}\!:\:\: 	&  \underset{{\mathbf{T}}} {\text{min}} \quad \frac{1}{2\mu}(||\mathbf{T}||_*-||\mathbf{T}||_2), \quad\text{s.t.} \:\:\eqref{12b}-\eqref{kkk},
		\end{align}
	\end{subequations}
where $\mu$ is the penalty factor \cite{Shayan_Zargari_1}. Specifically, for a sufficiently small value of $\mu$, solving (P6) yields a rank-one solution. However, (P6) is not convex as its objective function is the difference of two convex functions. To handle this, we resort to the SCA technique by approximating $\Psi(\mathbf{T})=||\mathbf{T}||_2$ with its first-order Taylor series expansion, which is a global lower bound as $\Psi(\mathbf{T})$ is convex. The first-order Taylor series  leads to 
	\begin{align}\label{sca}
	{\small \Psi(\mathbf{T})\geq \Psi(\mathbf{T}^i)+\text{Tr}\bigg(\nabla_{\mathbf{T}}^H\Psi(\mathbf{T}^{i})(\mathbf{T}-\mathbf{T}^{i})\bigg)\triangleq\tilde{\Psi}(\mathbf{T})},
	\end{align}
	where $
	\nabla_{\mathbf{T}} ~\|\mathbf{T}^{i}\|_2=\nabla_{\mathbf{T}} \bu_1^H\bQ^{i}\bu_1=\nabla_{\mathbf{T}} \text{Tr}\left(\bQ^{i}\bu_1\bu_1^H\right)=\bu_1\bu_1^H
	$, and $\bu_1$ is the eigenvector corresponding to the largest eigenvalue of $\bQ^{i}$.
	\begin{algorithm}[t]
		\caption{ {Overall   BCD  Algorithm}}
		\begin{algorithmic}[1]\label{algorithm_AO}
			\renewcommand{\algorithmicrequire}{\textbf{Input:}}
			\renewcommand{\algorithmicensure}{\textbf{Output:}}
			\REQUIRE Set the iteration number $i=0$, {the convergence tolerance $\zeta$}, and initialize  $\mathbf{T}=\mathbf{T}^{(0)}$.\\    
			\STATE \textbf{repeat}\\
			\STATE \quad Solve  (P3) for given $\mathbf{T}^{(i)}$ to obtain   $\{\mathbf{W}_k^{(i)},\rho_k^{(i)}\}$.
			\STATE \quad Solve  (P7) for given $\{\mathbf{W}_k^{(i)},\rho_k^{(i)}\}$ and   denote the \\ \quad solution as $\mathbf{T}^{(i+1)}$ based on  Algorithm 1. 
			\STATE \quad Set $i\leftarrow i+1$;
			\STATE  { \textbf{until} $\frac{|f_1^{(i)}-f_1^{(i-1)}|}{f_1^{(i-1)}}\geq \zeta$}
			\ENSURE return solution $\{\mathbf{T}^\ast\}$.
		\end{algorithmic}
	\end{algorithm}
	 While (P6) contains an objective function, it is still  a feasibility problem as any $\mathbf{T}$ that has rank-one and satisfies the constraints is an optimal solution. Thus, we enforce an optimal solution that optimizes SINR and harvested energy margins while satisfying the rank-one constraint. To achieve this goal, we propose two new slack variables, $\tau_k$  and $\Delta_k$ as the “SINR residual” and “EH residual,” respectively \cite{Zargari1}. The new formulated optimization problem is given by 
	 		\begin{subequations}
		\begin{align}
\hspace{-2mm}	\text{(P7)}\!:\:\:  \! 	& \underset{\mathbf{T},\tau_k ,\Delta_k} {\text{minimize}} \quad f_3 = \frac{1}{2\mu}(||\mathbf{T}||_*\!-\!\tilde{\Psi}(\mathbf{T}))\!-\!\sum_{k=1}^{K} \left(\alpha\tau_k+\beta \Delta_k\right),\\
		&\text{s.t.} \:\:  \text{Modified–}\eqref{12b},\: \text{Modified–}\eqref{12c},\: \eqref{kkk},\\
		&\:\:\quad  \tau_k,~\Delta_k\geq 0,~\forall k,
		\end{align}
	\end{subequations}
% 	\begin{subequations}
% 		\begin{align}
% 		\text{(P6)}:  \quad	&\underset{{\mathbf{T}}} {\text{minimize}} \:\: ||\mathbf{T}||_*-\tilde{\Psi}(\mathbf{T}),\\
% 		&\text{s.t.} \quad \eqref{12b}-\eqref{kkk}.~\label{31b}
% 		\end{align}
% 	\end{subequations}
% 	\begin{subequations}
% 		\begin{align}
% 		\text{(P7)}: \quad	& \underset{\mathbf{T},\tau_k ,\Delta_k} {\text{minimize}} \quad ||\mathbf{T}||_*-\tilde{\Psi}(\mathbf{T})-\sum_{k=1}^{K} \left(\alpha\tau_k+\beta \Delta_k\right),\\
% 		&\text{s.t.} \quad \frac{{\text{Tr}(\hat{\mathbf{A}}_{k,k})}}{{{\gamma _k}}} - \sum\limits_{\scriptstyle i = 1\atop\scriptstyle i \ne k}^K {\text{Tr}(\hat{\mathbf{A}}_{i,k})} \ge \sigma _k^2+ \frac{{\delta _k^2}}{{\rho _k }}+\tau_k,\\
% 		&\quad\quad \sum\limits_{i = 1}^K {{ \text{Tr}(\hat{\mathbf{A}}_{i,k}) }+\text{Tr}(\hat{\mathbf{B}}_k) \ge \frac{{{P_k^{\text{L}}(e_{k})}}}{{1 - {\rho_k}}}} +\Delta_k,~\forall k ,\\
% 		&\quad\quad \text{(\ref{kkk})},~\tau_k,~\Delta_k\geq 0,~\forall k ,
% 		\end{align}
% 	\end{subequations}
where {Modified–}\eqref{12b} and {Modified–}\eqref{12c}  are obtained
from \eqref{12b} and \eqref{12c} by substituting $\gamma_k$ with $\gamma_k+\tau_k$ and
${{P_k^{\text{L}}(e_{k})}}$ with ${{P_k^{\text{L}}(e_{k})}}+\Delta_k$, $\forall k\in \mathcal{K}$, respectively.  Besides, $\alpha$ and  $\beta$ are positive constants. The feasible set for both problems (P6) and (P7) is the same. However, (P7)  is more efficiently convergent \cite{Shayan_Zargari_1}. The IPPA and the overall BCD algorithm are summarized  in Algorithm 1 and  Algorithm 2, respectively. The order of complexity for solving (P7) is given by $\mathcal{O}(  \log({1}/{\epsilon})((3K+1)(N^{3.5}+3KN^{2.5})))$, where $\epsilon>0$ is the solution accuracy indicating that the active RIS does not have a higher computational complexity than the passive RIS \cite{Shayan_Zargari_1}. {Algorithm 2 iterations yield a non-increasing sequence of objective values for (P2) with guaranteed convergence. Let us consider $\{\mathbf{w}_k, \rho_k\}$ and ${\boldsymbol{\theta}}$ as the optimal solution of (P3) and (P7), respectively. Then, we have$ f_1({\boldsymbol{\theta}}^{(i+1)},\mathbf{w}^{(i+1)}_k, \rho^{(i+1)}_k )  =   f_1({\boldsymbol{\theta}}^{(i)},\mathbf{w}^{(i+1)}_k, \rho^{(i+1)}_k )  \leq  f_1({\boldsymbol{\theta}}^{(i)},\mathbf{w}^{(i)}_k, \rho^{(i)}_k ).$
Equality arises as the objective is not a function of ${\boldsymbol{\theta}}$.
So, as long as a feasible ${\boldsymbol{\theta}}$ is selected, equality remains valid. Inequality comes from the fact that for given ${\boldsymbol{\theta}}$, the solutions $\{\mathbf{w}_k, \rho_k\}$  are optimal. As for the convergence, we have a sequence of non-increasing objective values, which is guaranteed to converge.}

%The SCA-based algorithm of (P7) and the overall BCD algorithm are summarized in Algorithm 1 and  Algorithm 2, respectively. According to \cite[Theorem 1]{Shayan_Zargari_2}, this BCD   is guaranteed to generate a non-increasing sequence of objective function values. The order of complexity for solving (P7) is given by $\mathcal{O}(  \log({1}/{\epsilon})((3K+1)(N^{3.5}+3KN^{2.5})))$, where $\epsilon>0$ is the solution accuracy indicating that the active RIS does not have a higher computational complexity than the passive RIS \cite{Shayan_Zargari_1}.
		\begin{figure} 
			\centering
			\includegraphics[width=2.7in]{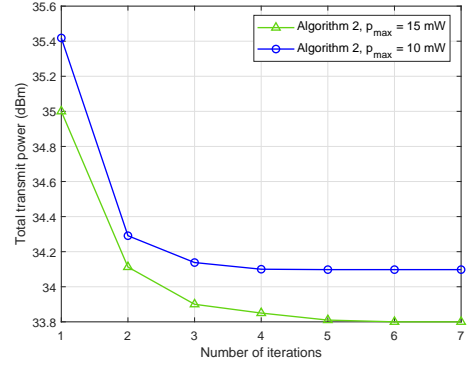}
			\caption{{Convergence  of Algorithm 2 ($N=20$ and  $M=10$).} \vspace{-5mm}}\label{fig_cov}
		\end{figure}

\begin{figure*}[t]
\centering
\begin{subfigure}[b]{0.32\textwidth}
\includegraphics[width=\textwidth]{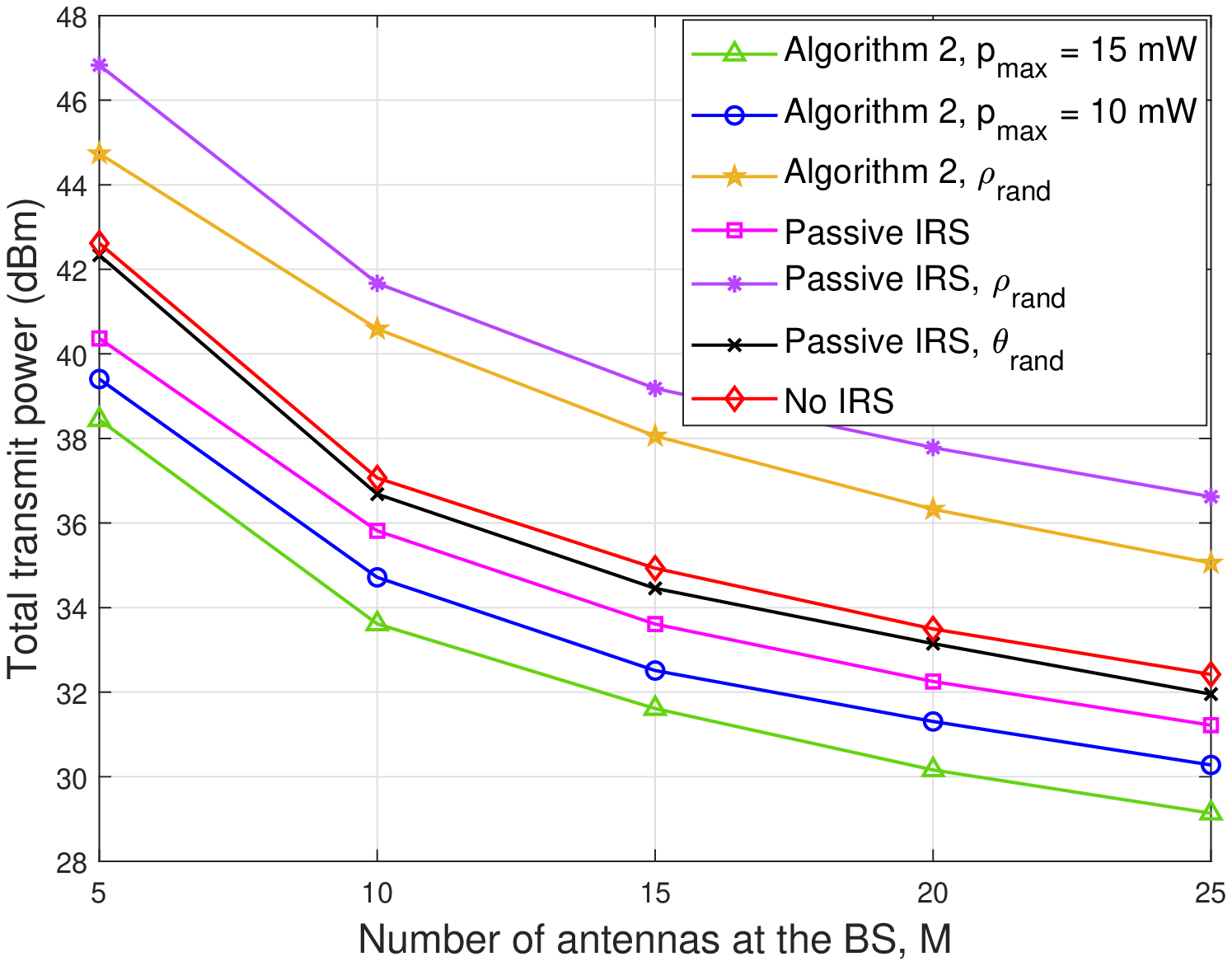}
\caption{ }   \label{antenna_fig}
\end{subfigure}
% \begin{subfigure}[b]{0.32\textwidth}
% \includegraphics[width=\textwidth]{fig4.eps}
% \caption{ }   \label{antenna_fig_2}
% \end{subfigure}
\begin{subfigure}[b]{0.32\textwidth}
\includegraphics[width=\textwidth]{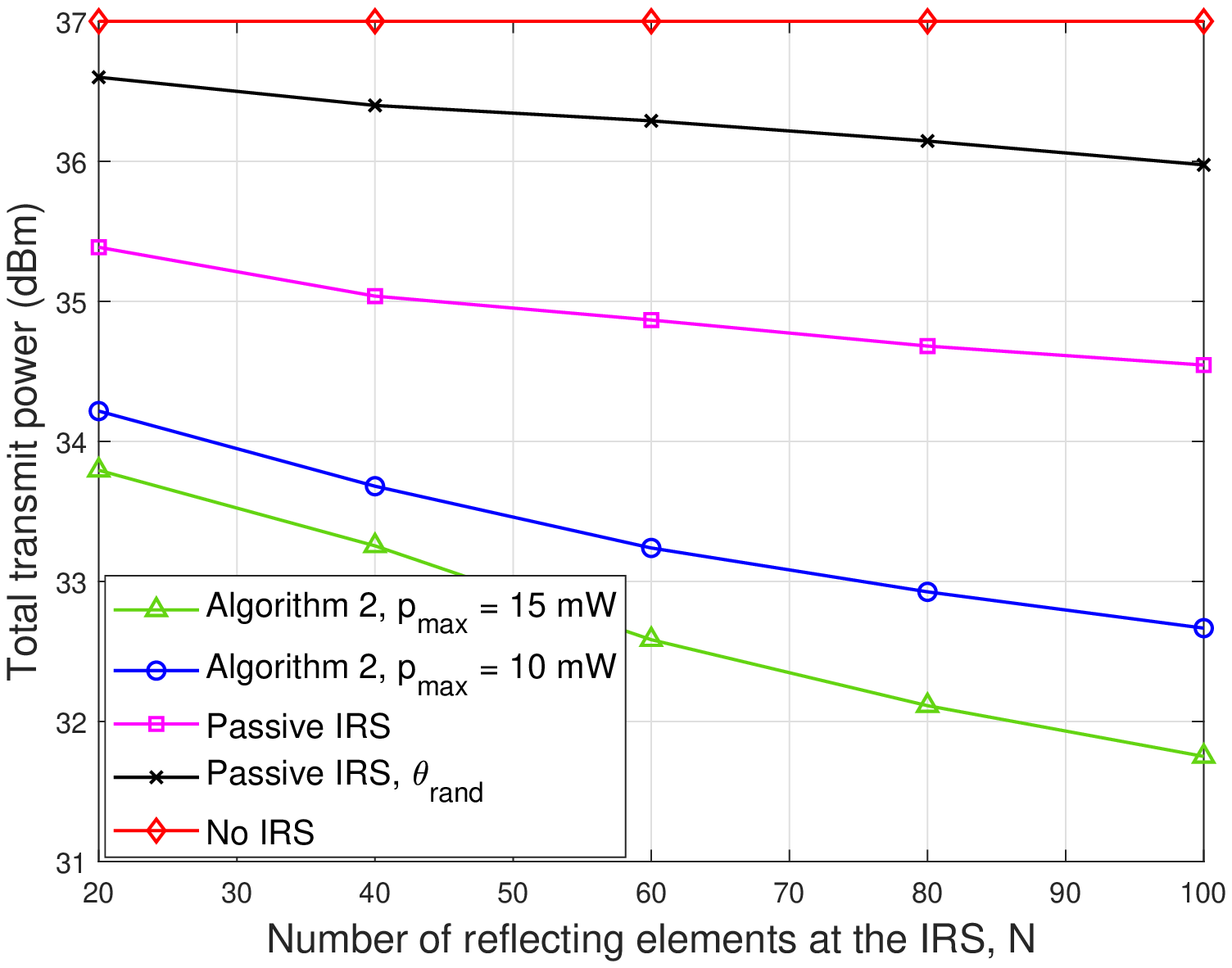}
\caption{ }      \label{phase_fig}
\end{subfigure}
\begin{subfigure}[b]{0.32\textwidth}
\includegraphics[width=\textwidth]{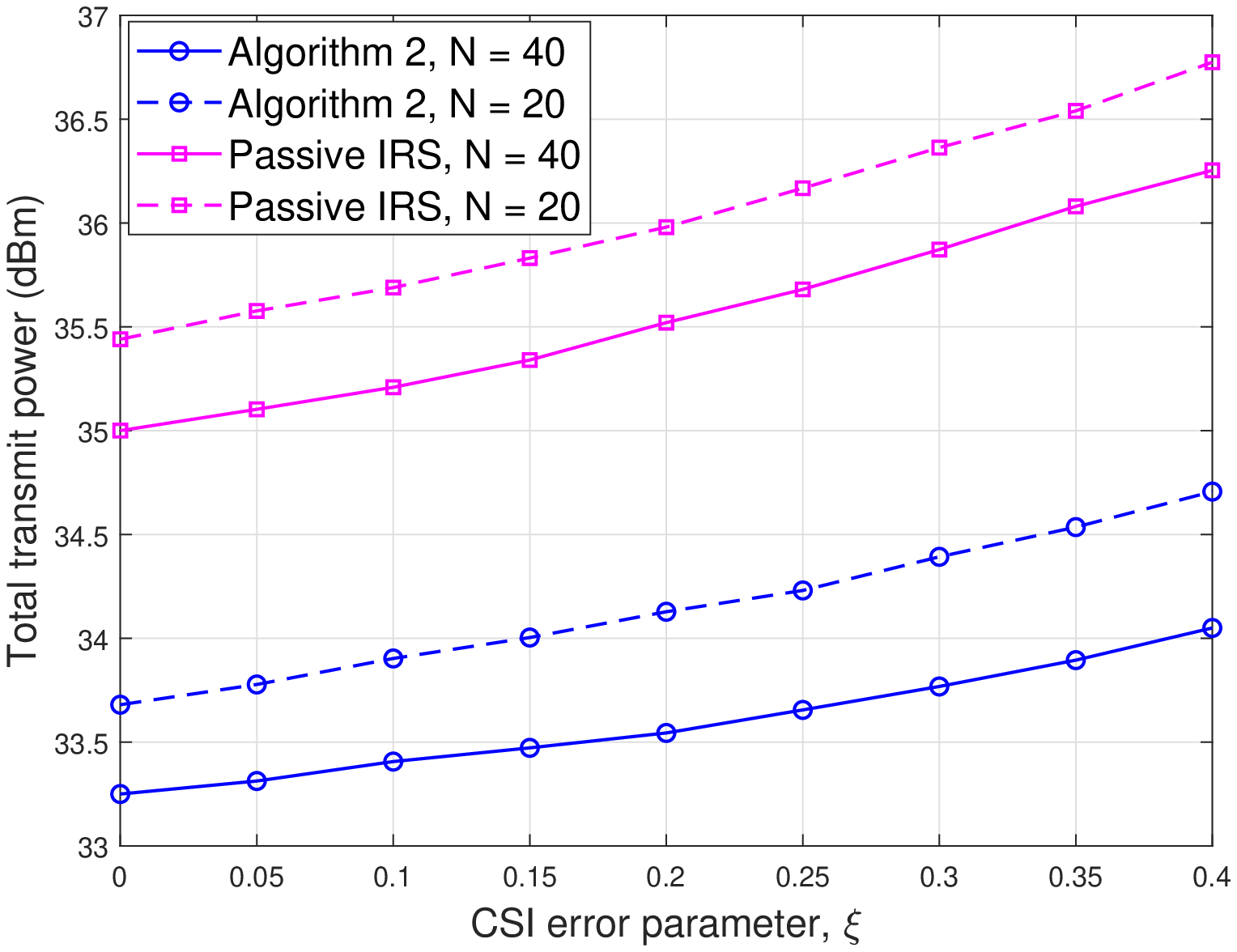}
\caption{ }   \label{channel_unc_fig}
\end{subfigure}
 \caption{ (\ref{antenna_fig})  show the transmit power versus number of antenna at the BS, $M$,  with $N=20$. (\ref{phase_fig}) shows the transmit power versus number of active elements at the RIS, $N$, with $M = 10$.  (\ref{channel_unc_fig}) represents the transmit power versus CSI error parameter, $\xi$, with $M=10$ and $p_{\max} = 15$ mW. \vspace{-5mm}}
\end{figure*}

\section{Simulation Results}
Here are the parameters of simulation results to  evaluate   {Algorithm \ref{algorithm_AO}}. The RIS has $N$ active elements. All users are randomly located in a cluster with  center $(3.5, 8)$ meters (m) and radius $2.5$ m. The AP location and the RIS location are considered as $(3.5, 0)$ m and $(0, 8)$ m, respectively. The Rician factor is set to $10$ dB, $K=4$, $\gamma_{k} = 10$ dB, $p_{\max}= 10$ mW, $e_k = -20$ dBm, $\sigma^2_k= -70$ dBm, $\delta_k^2 = -50$ dBm,   $\sigma^2_v =-70$  dBm, ${\zeta=10^{-3}}$, and $\mu=5\times 10^{-5}$ \cite{Shayan_Zargari_1,Shayan_Zargari_2}. The nonlinear EH model parameters are set to $a=2.463$, $b=1.635$, and $c=0.826$ \cite{Yunfei_Chen}. We consider the  pathloss model $L(d) = C_0(\frac{d}{D_0})^{-\kappa}$, where $C_0=-30$ dB denotes the pathloss at the reference distance $D_0 = 1$ m, $d$ indicates the link distance and $\kappa$ is the pathloss exponent, which is set to $3$ and $2.2$ for the direct and reflected links, respectively. For comparison, three benchmark system designs are studied, namely, 1) Algorithm $2$ with different maximum reflect power; 2) Algorithm $2$ with random PS ratio $(\rho_{\text{rnd}})$;  3) Passive RIS;  4) Passive RIS with $\rho_{\text{rnd}}$;  5) Passive RIS with random phase shifts $(\theta_{\text{rnd}})$;  6) No RIS.  For a fair comparison, the figures report the total transmit power as the sum of the  BS and active RIS powers.
%For a fair comparison, the figures report the total transmit power; in the passive RIS set up, it is the BS transmit power, and in the active RIS set up, it is the sum of the  BS and RIS powers. 

{Fig. \ref{fig_cov} shows the convergence behavior of Algorithm 2 for different maximum reflected power of the RIS,  $p_{\max}$. All the curves converge to a stationary point within less than five iterations on average.} Fig. (\ref{antenna_fig})  plots the total transmit power versus the number of BS antennas, $M$, for different benchmarks. When $M$ increases, the transmit power decreases for all the schemes. As well,  Algorithm 2 outperforms others in terms of transmit power and achieves better performance by allocating more power to the active RIS. Thus, employing it further reduces the total transmit power given an additional maximum reflected power. However, it is significant to optimize  $\rho$ because   $\rho_{\text{rnd}}$ increases the transmit power.  

Fig. (\ref{phase_fig}) represents the impact of the number of elements, $N$, at the active RIS on the total transmit power. As expected, sufficient reflectors at the active RIS cause transmit power reduction. When $N$ grows, the multipath propagation between the BS and the users increases, leading to the enhancement of the received signal power at the user. Thus, to get the minimum required SINR for users, the power at the BS  decreases. Indeed, with active RIS, the transmit power is reduced more due to the signal amplification at the RIS. This fact reveals that using the large or even massive elements at the active RIS can be advantageous for the MISO PS-SWIPT systems. Also,  with $N=100$, the active RIS with a maximum reflect power of $10$ mW can save the transmit power $ 19 \% $ compared to the passive RIS. For the active RIS with a maximum reflect power of $15$ mW, the gain is $ 28 \% $. 

Obtaining CSI    is a critical challenge for wireless networks, and imperfect CSI is normal. 
Thus,  it is important to study  the robustness of Algorithm 2  to CSI error. For that, we use the following channel estimation model \cite{Shayan_Zargari_2}:
$ \hat{h}=h+e$, where $h$ is the real channel and $e$ is the estimation error with Gaussian distribution and zero mean, i.e., $e \sim  \mathcal{N}(0,\,\sigma_{e}^2)$. We assume that the error variance satisfies $\sigma_{e}^2 \triangleq \xi |h|^2$, where $\xi$ denotes the level of CSI error.
Fig. (\ref{channel_unc_fig}) shows the total transmit power versus $\xi$, indicating that the transmit power grows by increasing $\xi$. Specifically, for the active RIS scheme with $N=40$ and CSI error power of $10\%$ of the channel gain (i.e., $\xi = 0.1$), the performance of Algorithm 2  experiences a loss of $5\%$  compared to the ideal case without CSI error (i.e., $\xi = 0$). But, it exhibits strong robustness against  CSI error.
 
\section{Conclusions}
	This letter studied the transmit power minimization problem for a  multiuser active RIS PS-SWIPT system. The BCD algorithm was developed to optimize the BS beamformers, RIS phase shifts/amplification factors, and PS ratios jointly. Minimum requirements of data rate and harvested energy based on the piecewise nonlinear EH model were considered. We established the convergence of Algorithm 2 to a locally optimal solution. Simulation results demonstrated that the submitted designs perform better than the benchmarks. We found that active RIS helps to overcome the double pathloss effect and improve the EH capability of the users. It does so while decreasing the total transmit power.  This letter serves as guidelines for many potential works on active RIS PS-SWIPT systems, e.g., prototype development, channel estimation, EH at the RIS, hardware design, and others.

\bibliographystyle{ieeetr}
\bibliography{ref}	

\begin{thebibliography}{10}

\bibitem{renzo2020}
M.~D. Renzo, A.~Zappone, M.~Debbah, M.~Alouini, C.~Yuen, J.~de~Rosny, and S.~A.
  Tretyakov, ``Smart radio environments empowered by reconfigurable intelligent
  surfaces: How it works, state of research, and the road ahead,'' {\em {IEEE}
  J. Sel. Areas Commun.}, vol.~38, pp.~2450--2525, 2020.

\bibitem{zhang2021}
Z.~Zhang, L.~Dai, X.~Chen, C.~Liu, F.~Yang, R.~Schober, and H.~V. Poor,
  ``Active {RIS} vs. passive {RIS:} which will prevail in {6G?},'' {\em CoRR},
  vol.~abs/2103.15154, 2021.

\bibitem{Shaokang}
S.~Hu, Z.~Wei, Y.~Cai, C.~Liu, D.~W.~K. Ng, and J.~Yuan, ``Robust and secure
  sum-rate maximization for multiuser {MISO} downlink systems with
  self-sustainable irs,'' {\em IEEE Trans. Commun.}, vol.~69, no.~10,
  pp.~7032--7049, 2021.

\bibitem{long2021}
R.~Long, Y.~Liang, Y.~Pei, and E.~G. Larsson, ``Active reconfigurable
  intelligent surface-aided wireless communications,'' {\em {IEEE} Trans.
  Wirel. Commun.}, vol.~20, no.~8, pp.~4962--4975, 2021.

\bibitem{xu2021}
D.~Xu, X.~Yu, D.~W. Kwan~Ng, and R.~Schober, ``Resource allocation for active
  {IRS}-assisted multiuser communication systems,'' in {\em 55th Asilomar
  Conference on Signals, Systems, and Computers}, pp.~113--119, 2021.

\bibitem{YouZ21}
C.~You and R.~Zhang, ``Wireless communication aided by intelligent reflecting
  surface: Active or passive?,'' {\em {IEEE} Wirel. Commun. Lett.}, vol.~10,
  no.~12, pp.~2659--2663, 2021.

\bibitem{JungSK21}
M.~Jung, W.~Saad, and G.~Kong, ``Performance analysis of active large
  intelligent surfaces ({LISs}): Uplink spectral efficiency and pilot
  training,'' {\em {IEEE} Trans. Commun.}, vol.~69, no.~5, pp.~3379--3394,
  2021.

\bibitem{Zargari1}
S.~Zargari, S.~Farahmand, and B.~Abolhassani, ``Joint design of transmit
  beamforming, {IRS} platform, and power splitting {SWIPT} receivers for
  downlink cellular multiuser {MISO},'' {\em Phys. Commun.}, vol.~48, 2021.

\bibitem{Shayan_Zargari_1}
S.~Zargari, A.~Khalili, Q.~Wu, M.~R. Mili, and D.~W.~K. Ng, ``Max-min fair
  energy-efficient beamforming design for intelligent reflecting surface-aided
  {SWIPT} systems with non-linear energy harvesting model,'' {\em {IEEE} Trans.
  Veh. Technol.}, vol.~70, no.~6, pp.~5848--5864, 2021.

\bibitem{Shayan_Zargari_2}
S.~Zargari, S.~Farahmand, B.~Abolhassani, and C.~Tellambura, ``Robust active
  and passive beamformer design for {IRS}-aided downlink {MISO} {PS-SWIPT} with
  a nonlinear energy harvesting model,'' {\em {IEEE} Trans. Green Commun.
  Netw.}, vol.~5, no.~4, pp.~2027--2041, 2021.

\bibitem{Yunfei_Chen}
Y.~Chen, N.~Zhao, and M.~Alouini, ``Wireless energy harvesting using signals
  from multiple fading channels,'' {\em {IEEE} Trans. Commun.}, vol.~65,
  no.~11, pp.~5027--5039, 2017.

\bibitem{18}
M.~Grant, S.~Boyd, and Y.~Ye, ``{CVX}: Matlab software for disciplined convex
  programming,'' 2009.

\end{thebibliography}
\end{document}